\documentclass[preprint,onecolumn,nofootinbib]{revtex4}
\pdfoutput=1

\usepackage[colorlinks=true,linkcolor=blue,urlcolor=blue,filecolor=black,citecolor=red,pdfstartview=FitV,pdftitle={},pdfsubject={},pdfkeywords={},pdfpagemode=None,bookmarksopen=true]{hyperref}
\usepackage{graphicx}
\usepackage{amsmath}
\usepackage{amsfonts}
\usepackage{amssymb,ulem}
\usepackage{color}%
\usepackage{dcolumn}
\newcommand{\f}{\begin{equation}}
\newcommand{\ff}{\end{equation}}
\newcommand{\fa}{\begin{eqnarray}}
\newcommand{\ffa}{\end{eqnarray}}

\begin{document}
\title{Holographic quantum critical conductivity from higher derivative electrodynamics}
\author{Jian-Pin Wu}
\thanks{jianpinwu@yzu.edu.cn}
\affiliation{
$^1$Center for Gravitation and Cosmology, College of Physical Science and Technology,
Yangzhou University, Yangzhou 225009, China\ \\
$^2$ Institute of Gravitation and Cosmology, Department of
Physics, School of Mathematics and Physics, Bohai University, Jinzhou 121013, China}
\begin{abstract}

We study the conductivity from higher derivative electrodynamics in a holographic quantum critical phase (QCP).
Two key features of this model are observed.
First, a rescaling for the Euclidean frequency by a constant is needed when fitting the quantum Monte Carlo (QMC) data for the $O(2)$ QCP.
We conclude that it is a common characteristic of the higher derivative electrodynamics.
Second, both the Drude-like peak at low frequency and the pronounced peak can simultaneously emerge.
They are more evident for the relevant operators than for the irrelevant operators.
In addition, our result also further confirms that the conductivity for the $O(2)$ QCP is particle-like but not vortex-like.
Finally, the electromagnetic (EM) duality is briefly discussed.
The largest discrepancies of the particle-vortex duality in the boundary theory appear at the low frequency
and the particle-vortex duality holds more well for the irrelevant operator than for the relevant operator.

\end{abstract}
\maketitle
\section{Introduction}

Quantum critical (QC) system, which includes quantum critical phase transition (QCPT) and quantum critical phase (QCP),
is a long-standing important issue in condensed matter physics \cite{Sachdev:2000qpt}.
Some of the best understood examples are described by strongly interacting conformal field theory (CFT) at low energy.
A canonical example is the superfluid-insulator QCP described by the boson Hubbard model.

The real-time dynamics, especially the frequency-dependent conductivity $\sigma(\omega)$, at finite temperature is a central and challenging subject in QC physics \cite{Damle:1997rxu}.
Because of the strongly correlated nature of QC system,
the conventional perturbative methods in traditional quantum field theory (QFT) unfortunately lose its power in studying the dynamics.
The novel non-perturbative techniques and methods are called for.

AdS/CFT correspondence \cite{Maldacena:1997re,Gubser:1998bc,Witten:1998qj,Aharony:1999ti}
provides a powerful tool in dealing with the real-time dynamics of the strongly interacting QC system lacking quasi-particles.
References \cite{Witczak-Krempa:2013nua,Witczak-Krempa:2013aea,Katz:2014rla} construct holographic models based on
the Maxwell-Weyl system in Schwarzschild-AdS (SS-AdS) to study QC physics, in particular the dynamical conductivity $\sigma(\omega)$.
By combining high precision quantum Monte Carlo (QMC) results \cite{Katz:2014rla,Witczak-Krempa:2013nua,Chen:2013ppa}
for the dynamical conductivity in the $O(2)$ QCP with that from holography \cite{Witczak-Krempa:2013nua,Witczak-Krempa:2013aea,Katz:2014rla},
they build a quantitative description of the dynamics of QC systems lacking quasi-particles
and find that the dynamical conductivity for the $O(2)$ QCP is particle-like but not vortex-like,
which resolved the puzzle of $O(2)$ QCP.

Further studies find that the relevant scalar operator plays a key role in the dynamics of the QC systems \cite{Katz:2014rla}.
However, the scalar field in the bulk introduced in \cite{Katz:2014rla}, which is dual to the relevant operator in the boundary field theory,
is not a dynamical field. To overcome this shortcoming, references \cite{Myers:2016wsu,Lucas:2017dqa} construct a novel neutral scalar hair black brane
by coupling Weyl tensor with neutral scalar field, which provides a framework to describe QC dynamics and the one away from QCP.
In particular, the relevant operator of this model acquires a thermal expectation value
and we can study the dynamical conductivity for a wide range of conformal dimensions $\Delta$.

However, until now AdS/CFT is best understood only at large-N limit \cite{Maldacena:1997re,Gubser:1998bc,Witten:1998qj,Aharony:1999ti}.
Therefore, it is important to study the universality and the speciality of the dynamics of the QC systems.
To this end, here we extend the studies in \cite{Myers:2016wsu,Lucas:2017dqa} to include a higher derivative term by incorporating a interaction between gauge field and Weyl tensor
and explore the generic and special properties of the holographic QC dynamics.

\section{Holographic framework}

We start with the following SS-AdS black brane
\begin{subequations}
\label{bl-br}
\begin{align}
ds^2=&\frac{r_0^2}{L^2u^2}\Big(-f(u)dt^2+dx^2+dy^2\Big)+\frac{L^2}{u^2f(u)}du^2\,,
\
\\
f(u)=&(1-u)p(u)\,,~~~~~~~
p(u)=u^2+u+1\,.
\end{align}
\end{subequations}
$u=0$ is the asymptotically AdS boundary while the horizon locates at $u=1$.
The Hawking temperature of this system is
\fa
T=\frac{3r_0}{4\pi L^2}\,.
\ffa

We study the following bulk action including a massless gauge field $A_{\mu}$,
and a scalar field $\Phi$
\begin{subequations}
\label{action}
\begin{align}
\label{action-phi}
S_{\Phi}=&-\frac{1}{2l_p^2}\int d^4x\sqrt{-g}\Big[(\nabla_{\mu}\Phi)^2+m^2\Phi^2-2\alpha_1L^2\Phi C^2\Big]\,,
\
\\
\label{action-A}
S_A=&-\int d^4x\sqrt{-g}\Big(\frac{1}{8g_F^2}F_{\mu\nu}X^{\mu\nu\rho\sigma}F_{\rho\sigma}\Big)\,.
\end{align}
\end{subequations}
The scalar field $\Phi$ in bulk gravity is dual to the scalar operator $\mathcal{O}$ with conformal dimension $\Delta=\frac{1}{2}(3\pm\sqrt{9+4m^2L^2})$ in boundary CFT.
In the action $S_A$, $F=dA$ is the curvature of gauge field $A$ and the tensor $X$ is
\fa
X_{\mu\nu}^{\ \ \rho\sigma}&=&
(1+\alpha_2\Phi)I_{\mu\nu}^{\ \ \rho\sigma}-8\gamma\Phi L^4 C_{\mu\nu}^{\ \ \rho\sigma}
\,,
\label{X-tensor}
\ffa
where $I_{\mu\nu}^{\ \ \rho\sigma}=\delta_{\mu}^{\ \rho}\delta_{\nu}^{\ \sigma}-\delta_{\mu}^{\ \sigma}\delta_{\nu}^{\ \rho}$
is an identity matrix.
In the above equations (\ref{action}), we have introduced the factors of $l_p$ and $L$
so that the coupling parameters $g_F$, $\alpha_{1,2}$, $\gamma$, and the scalar field $\Phi$ are dimensionless.
Without loss of generality, we set $l_p^2=1/2$, $g_F=1$ and $L=1$ in what follows.
Comparing with \cite{Myers:2016wsu}, we introduce a new interaction term in Eq. \eqref{X-tensor}
that coupling among the Weyl tensor, gauge field and scalar field.

The black brane geometry \eqref{bl-br} describes a thermal state in the dual boundary CFT.
Following the strategy in \cite{Myers:2016wsu}, we introduce an interaction term between the scalar field $\Phi$ and the Weyl tensor
such that the scalar field have a nontrivial profile in the black brane background,
which corresponds to a nonvanishing thermal expectation value of scalar operator in boundary theory.

From the action \eqref{action}, we obtain the EOMs for the scalar field and gauge field as
\begin{subequations}
\label{eom}
\begin{align}
\label{eom-phi-v0}
&(\nabla^2-m^2)\Phi+\alpha_1L^2C^2
-\frac{1}{16}(\alpha_2I_{\mu\nu}^{\ \ \rho\sigma}-8\gamma C_{\mu\nu}^{\ \ \rho\sigma})F^{\mu\nu}F_{\rho\sigma}
=0\,,
\
\\
\label{eom-max}
&\nabla_{\nu}(X^{\mu\nu\rho\sigma}F_{\rho\sigma})=0\,.
\end{align}
\end{subequations}
Since here we consider a thermal state, which described by the neutral black brane background.
In this case, the background gauge field is zero.
Therefore, Eq.\eqref{eom-phi-v0} reduces to
\fa
\label{eom-phi}
(\nabla^2-m^2)\Phi+\alpha_1L^2C^2=0\,.
\ffa
The above EOM determines the profile of the scalar field.

Since the Weyl tensor vanishes in the AdS boundary,
the asymptotic behavior of $\Phi(u)$ is the same as that without the $\alpha_1$ coupling term, which behaves
\fa
\label{asy-beh-phi}
\Phi(u)=\Phi_0 u^{3-\Delta}+\Phi_1 u^{\Delta}\,.
\ffa
We identify $\Phi_0$ as the source, which corresponds to the coupling of the boundary QFT and deforms it,
and $\Phi_1$ as the expectation.
The conformal dimension $\Delta$ is constrained in $\Delta\geq 1/2$
such that the dual CFTs are unitary \cite{Klebanov:1999tb}.
When $\Phi_0=0$, the dual theory is the QCP \cite{Myers:2016wsu}.
If we tune $\Phi_0$ nonzero, the dual theory is away from QCP \cite{Myers:2016wsu}.
In this paper, we only focus on the case of $\Phi_0=0$.

\begin{figure}
\center{
\includegraphics[scale=0.65]{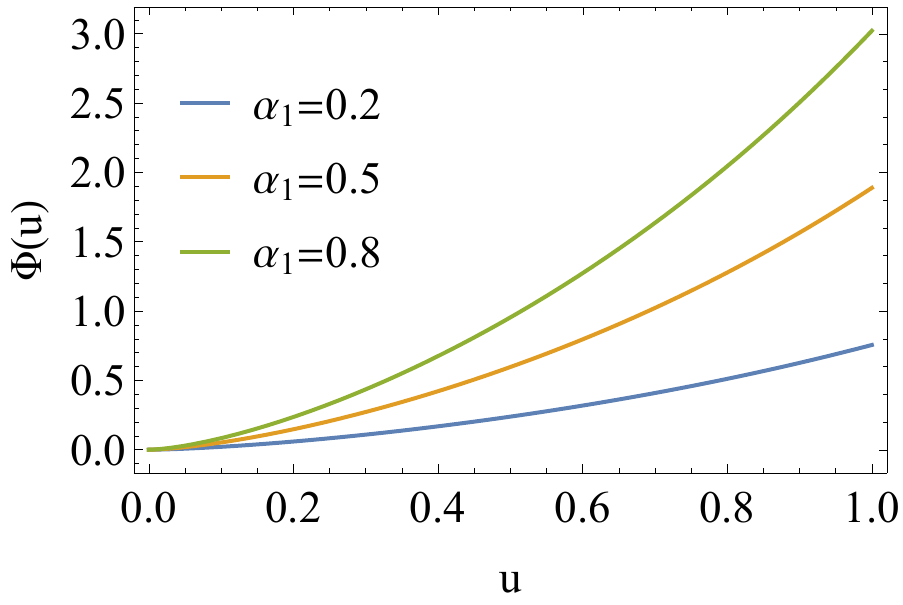}\ \hspace{0.8cm}
\includegraphics[scale=0.65]{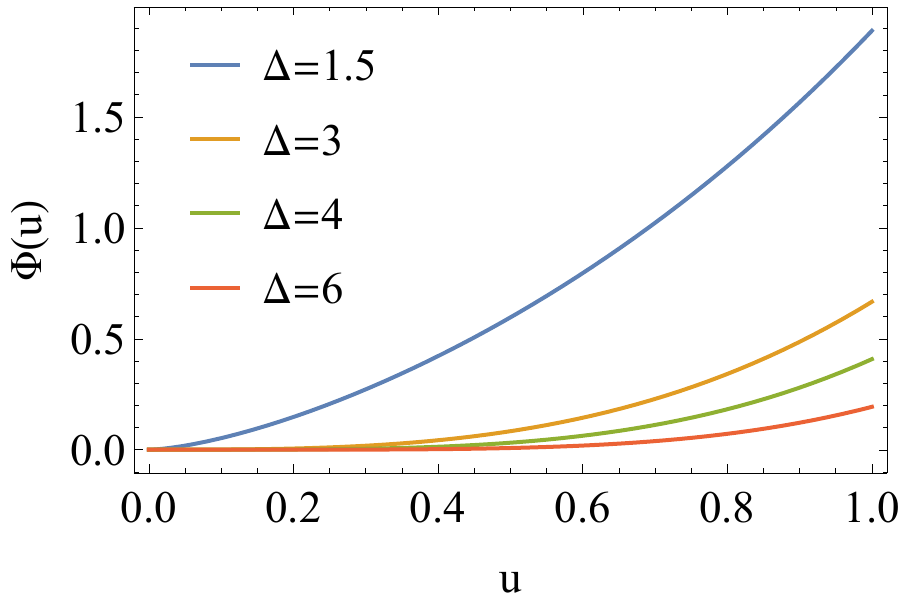}\ \hspace{0.8cm}\ \\
\caption{\label{phivsu} Plots of $\Phi(u)$ as the function of $u$ for sample $\Delta$ and $\alpha_1$.
Left plot is for $\Delta=1.5$ and different $\alpha_1$ and right plot is for $\alpha_1=0.5$ and different $\Delta$.
}}
\end{figure}

Combining the falling of $\Phi(u)$ (Eq. \eqref{asy-beh-phi}) at the boundary $u\rightarrow 0$ and the regular requirement of $\Phi(u)$ at the horizon,
we can numerically solve this EOM and show the profile of scalar field for sample $\Delta$ and $\alpha_1$ in FIG.\ref{phivsu}.
From this figure, we can see that the value of $\Phi(u)$ at the horizon increases with the increase of $\alpha_1$ for fixed $\Delta$.
While for fixed $\alpha_1$, the value of $\Phi(u)$ at the horizon increases with the decrease of $\Delta$.

\section{Holographic conductivity}

To calculate the frequency-dependent conductivity,
we turn on the perturbation of the gauge field at zero momentum along $y$ direction in Fourier space as $A_y\sim a_y(u)e^{-i\omega t}$
and the EOM for the gauge field \eqref{eom-max} can be explicitly wrote down as
\begin{align}
a_y''+a_y' \left(\frac{f'}{f}+\frac{3 \alpha _2 \Phi '-2 \gamma  u \left(f^{(3)} u
   \Phi +f'' \left(u \Phi '+2 \Phi \right)\right)}{3 \alpha _2 \Phi -2 \gamma  u^2 \Phi
   f''+3}\right)
   +\frac{\omega ^2 a_y}{f^2}
   =0\,.\label{eom-max-exp}
\end{align}
And then, the conductivity is given by
\fa
\label{con-def}
\sigma(\omega)=\frac{\partial_ua_y}{i\omega a_y}\Big |_{u\rightarrow 0}\,.
\ffa
Imposing the ingoing boundary condition at the horizon, we can numerically solve
the EOM \eqref{eom-max-exp} and read off the conductivity by Eq.\eqref{con-def}.

Subsequently, we shall study the conductivity at QCP by tuning $\Phi_0=0$.
We mainly study the properties of the conductivity from higher derivative electrodynamics in the holographic framework described in the last section.

\begin{figure}
\center{
\includegraphics[scale=0.5]{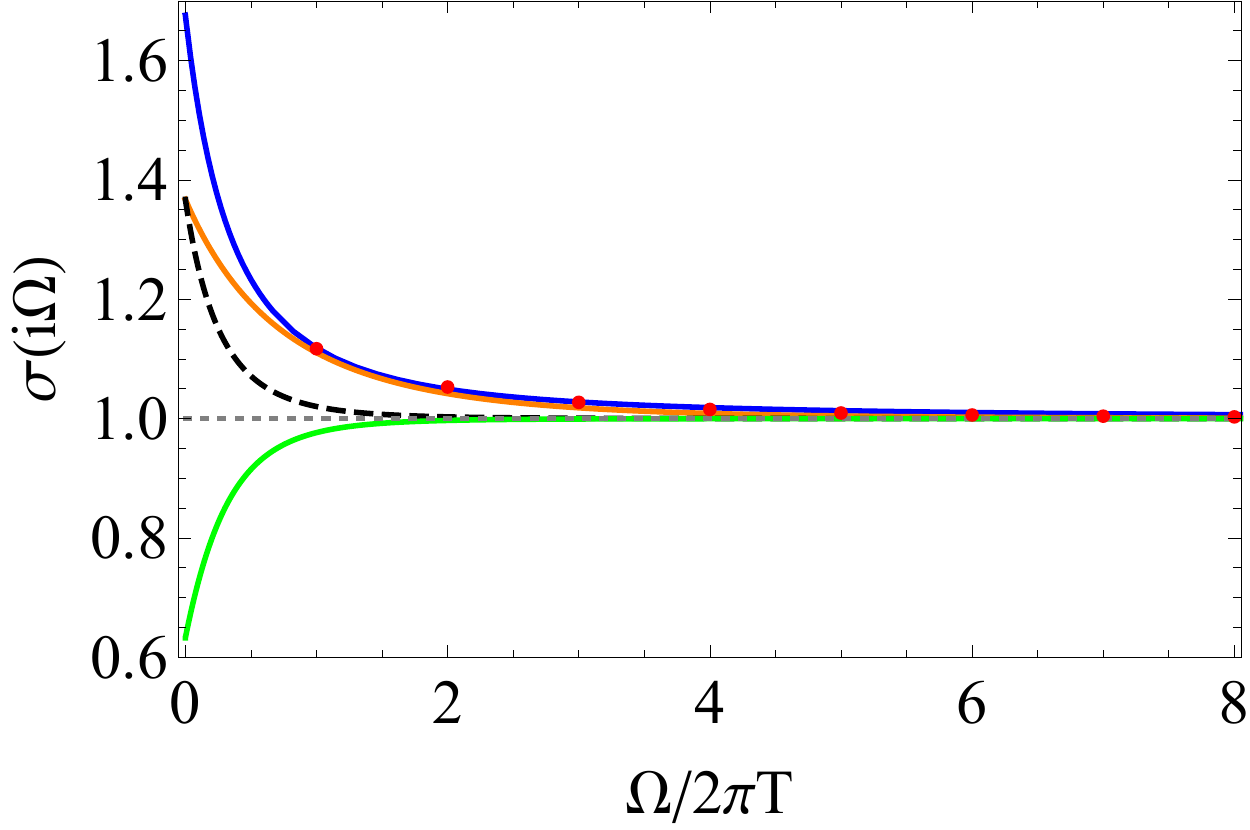}\ \hspace{0.8cm}
\includegraphics[scale=0.5]{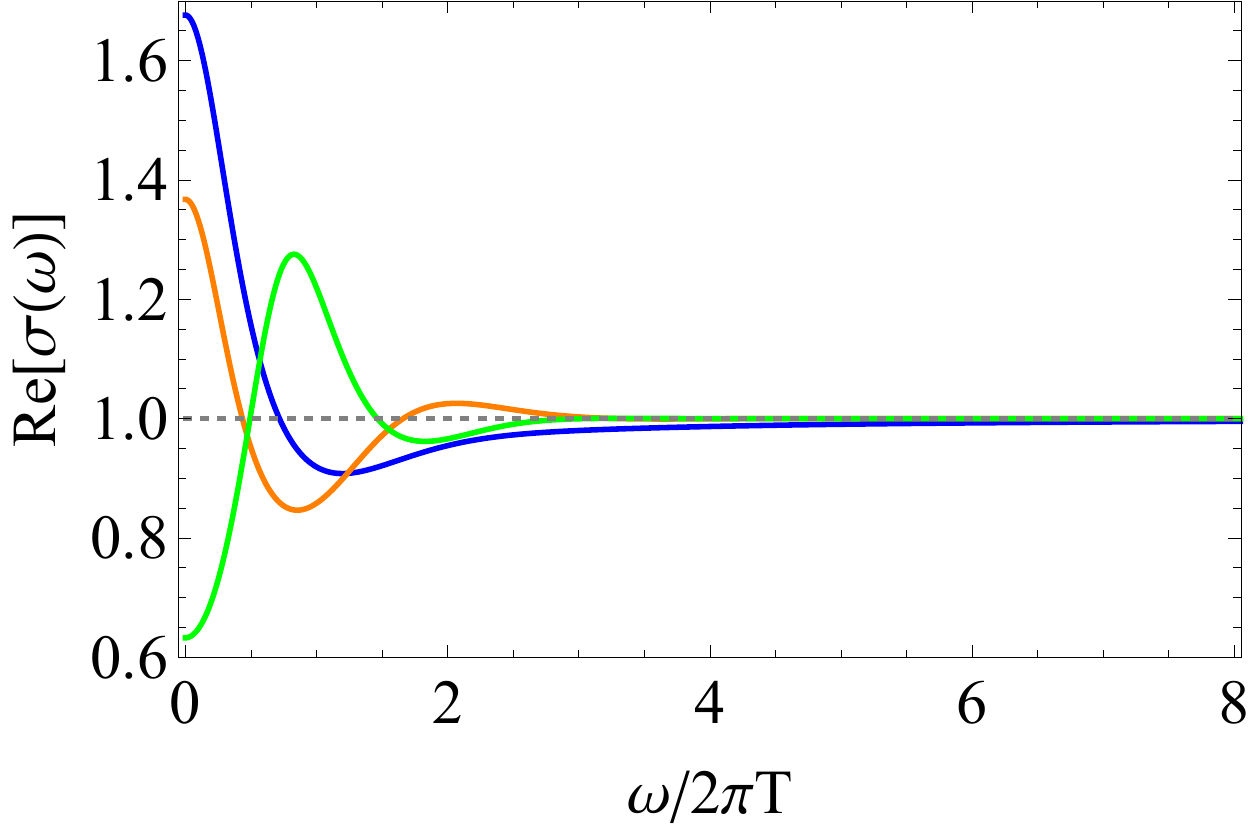}\ \hspace{0.8cm}\ \\
\caption{\label{sigma} Plots of the conductivity for Euclidean (left) and real (right) frequency for $\Delta=1.5$.
The red dots represent QMC data for the $O(2)$ QCP \cite{Katz:2014rla,Witczak-Krempa:2013nua,Chen:2013ppa}.
Blue lines are for $\alpha_2=0.589$ and $\gamma=0$. Orange and black dashed lines are for $\alpha_2=0$ and $\gamma=0.08$
(orange line in left plot with a rescaling of the $\Omega/T$ by $\alpha=0.35$ to fit the QMC data). Green lines are for $\alpha_2=0$ and $\gamma=-0.08$.
Here, $\alpha_1$ has been fixed as $\alpha_1=0.304$ such that we have the same scalar profile.
}}
\end{figure}

Previously, in \cite{Myers:2016wsu}, only when the $\alpha_2$ term survives, i.e., turning off $\gamma$ here,
it has been shown that the conductivity for Euclidean frequency can be fitted very well for $\Omega>2\pi T$
to the QMC data for the $O(2)$ QCP \cite{Katz:2014rla,Witczak-Krempa:2013nua,Chen:2013ppa}.
In fact, before that, the authors in \cite{Witczak-Krempa:2013nua} have found that the QMC data for the $O(2)$ QCP
can also be fitted in a simple holographic model in which only the coupling term between gauge field and Weyl tensor
is introduced in SS-AdS background. But in \cite{Witczak-Krempa:2013nua}, a rescaling for the Euclidean frequency $\Omega/T$ by a constant $\alpha=0.35$ is needed.

FIG.\ref{sigma} shows $\sigma(\omega)$ as a function of Euclidean and real frequency for $\Delta=1.5$.
We set $\alpha_1=0.304$ such that we have the same scalar profile
and can make comparison between $\alpha_2$ term and the higher derivative term $\gamma$.
We find that there is a common characteristic that to fit the QMC data, we need a rescaling for the Euclidean frequency $\Omega/T$ by the same constant $\alpha=0.35$
whether the nontrivial scalar profile is introduced or not.
It is different from the case of $\alpha_2$ term studied in \cite{Myers:2016wsu},
in which no rescaling is needed. It seems to indicate that there are some differences between the $O(2)$ QCP and the holographic dual CFTs from higher derivative electrodynamics.
Nevertheless, as has been illuminated in \cite{Witczak-Krempa:2013nua}, this rescaling does not affect the essential characteristics of conductivity.

We also plot the conductivity for $\gamma=-0.08$ and find that the conductivity for Euclidean frequency (green line) is inconsistent with the QCM data for $O(2)$ QCP.
It further confirms that the conductivity for the $O(2)$ QCP is particle-like but not vortex-like.
\begin{figure}
\center{
\includegraphics[scale=0.65]{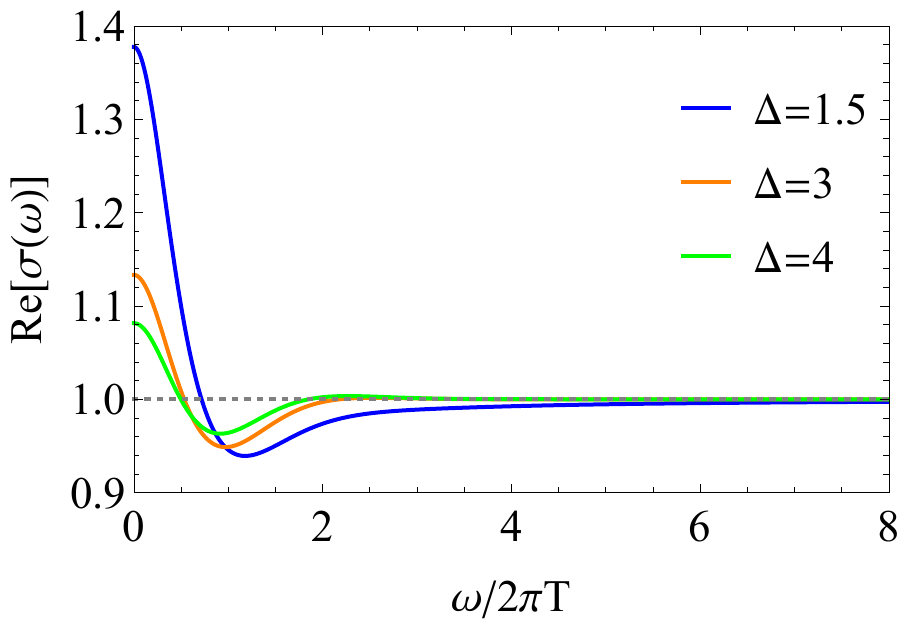}\ \hspace{0.8cm}
\includegraphics[scale=0.65]{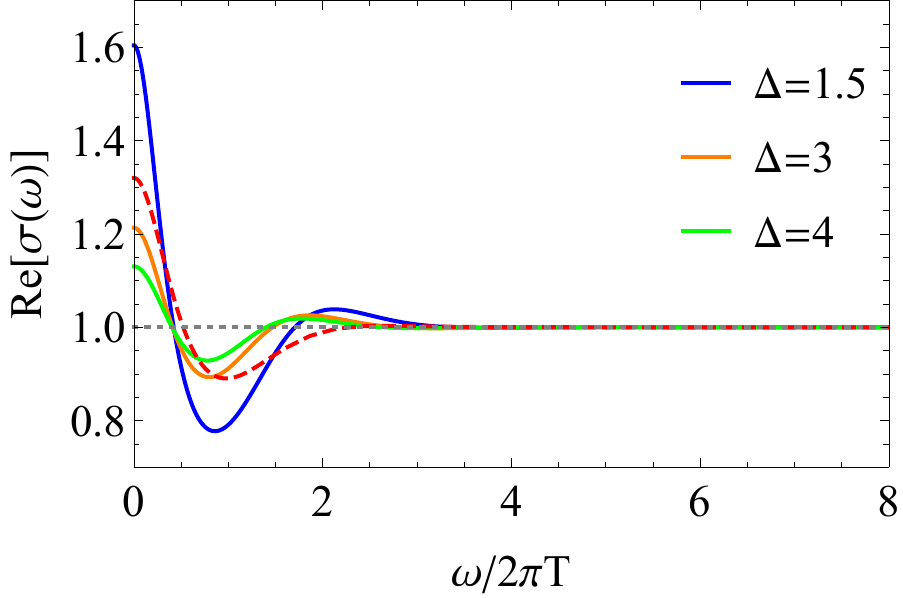}\ \hspace{0.8cm}\ \\
\caption{\label{sigma-diff-delta} Plots of the conductivity for real frequency for different $\Delta$.
Left plot is for $\alpha_2=0.2$, $\gamma=0$ and right plot for $\gamma=0.08$, $\alpha_2=0$.
Here we have set $\alpha_1=0.5$.
The red dashed line is for $\Phi=0$ with $\gamma=0.08$.
}}
\end{figure}

In addition, we also observe a novel property of our model that for $\gamma>0$, after a Drude-like peak appears at low frequency,
a small but evident pronounced peak emerges at intermediate frequency (FIG.\ref{sigma} and FIG.\ref{sigma-diff-delta}),
which is different from that in \cite{Myers:2016wsu,Witczak-Krempa:2013nua}.
In particular, both the Drude-like peak and the pronounced peak are more evident for the relevant operators ($\Delta<3$) than for irrelevant operators ($\Delta>3$).
It indicates that the emergence of the pronounced peak at intermediate frequency doesn't transfer from the Drude-like peak at the low frequency.
Anyhow, the pronounced peak is associated with the coupling between nontrivial scalar field dual to a relevant operator, and Weyl tensor.

\begin{figure}
\center{
\includegraphics[scale=0.65]{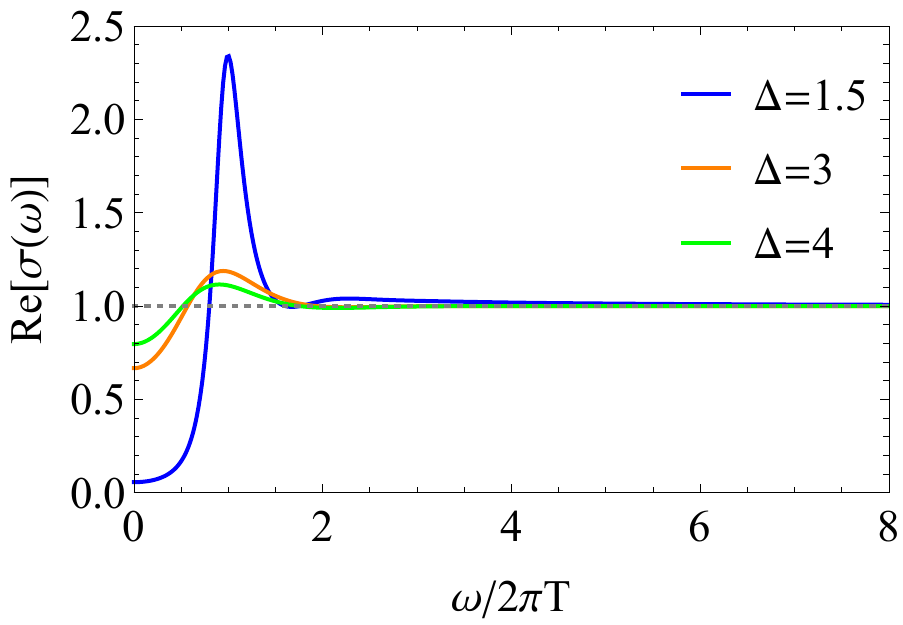}\ \hspace{0.8cm}
\includegraphics[scale=0.65]{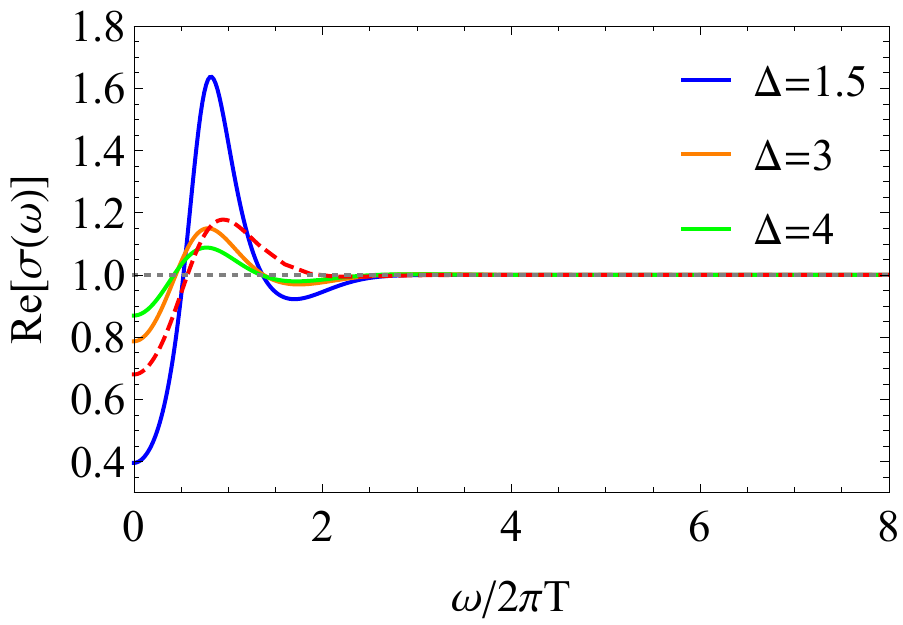}\ \hspace{0.8cm}\ \\
\caption{\label{sigma-nalpha2} Plots of the conductivity for real frequency with different $\Delta$.
Left plot is for $\alpha_2=-0.5$ and $\gamma=0$.
Right plot is for $\alpha_2=0$ and $\gamma=-0.08$.
Here we have set $\alpha_1=0.5$.
Note that the red dashed line in right plot is for $\Phi=0$ with $\gamma=-0.08$.
}}
\end{figure}

When the coupling parameters $\alpha_2$ and $\gamma$ are negative,
such pronounced peak at the intermediate frequency emerges whether there is the coupling of Weyl tensor or not (see FIG.\ref{sigma-nalpha2}).
Similar with the case of positive $\alpha_2$ and $\gamma$,
the pronounced peak are more evident for the relevant operators ($\Delta<3$) than for the irrelevant operators ($\Delta>3$).
But we stress that this pronounce peak for negative coupling parameters is different from the one for positive parameters.
Most of the spectral weight at the intermediate frequency for negative coupling parameters is from the transfer of the low frequency
but not for positive coupling parameters.
Note that in \cite{Lucas:2017dqa}, a pronounce peak at intermediate frequency is also observed in a system with finite charge density,
which is similar the case here for negative coupling parameters.

\section{EM duality}

In this section, we shall briefly discuss the electromagnetic (EM) duality.
Since the introduction of the coupling terms of $\alpha_2$ and $\gamma$,
the EM self-duality is violated.
But we can construct the corresponding EM dual theory as \cite{Myers:2010pk,Wu:2016jjd,Wu:2018pig,Wu:2018vlj},
which is
\fa
\label{ac-SB}
S_B=\int d^4x\sqrt{-g}\Big(-\frac{1}{8\hat{g}_F}G_{\mu\nu}\widehat{X}^{\mu\nu\rho\sigma}G_{\rho\sigma}\Big)\,,
\ffa
where $\hat{g}_F^2\equiv 1/g_F^2$ and $G_{\mu\nu}\equiv\partial_{\mu}B_{\nu}-\partial_{\nu}B_{\mu}$.
The tensor $\widehat{X}$ is given by
\fa
&&
\widehat{X}_{\mu\nu}^{\ \ \rho\sigma}=-\frac{1}{4}\varepsilon_{\mu\nu}^{\ \ \alpha\beta}(X^{-1})_{\alpha\beta}^{\ \ \gamma\lambda}\varepsilon_{\gamma\lambda}^{\ \ \rho\sigma}\,,
\label{X-hat}
\
\\
&&
\frac{1}{2}(X^{-1})_{\mu\nu}^{\ \ \rho\sigma}X_{\rho\sigma}^{\ \ \alpha\beta}\equiv I_{\mu\nu}^{\ \ \alpha\beta}\,,
\label{X-ne-def}
\ffa
where $\varepsilon_{\mu\nu\rho\sigma}$ is volume element.
And then, we can derive the EOM of the dual theory as
\fa
\nabla_{\nu}(\widehat{X}^{\mu\nu\rho\sigma}G_{\rho\sigma})=0\,.
\label{eom-Max-B}
\ffa

For the standard four-dimensional Maxwell theory, $\widehat{X}_{\mu\nu}^{\ \ \rho\sigma}=I_{\mu\nu}^{\ \ \rho\sigma}$ and thus the theory (\ref{action-A}) and (\ref{ac-SB})
are identical. It means that the Maxwell theory is self-dual.
When the coupling terms of $\alpha_2$ and $\gamma$ are introduced, we find that for small $\alpha_2$ and $\gamma$,
\fa
&&
(X^{-1})_{\mu\nu}^{\ \ \rho\sigma}=I_{\mu\nu}^{\ \ \rho\sigma}-\alpha_2\Phi I_{\mu\nu}^{\ \ \rho\sigma}+8\gamma\Phi C_{\mu\nu}^{\ \ \rho\sigma}+\mathcal{O}(\alpha_2^2)+\mathcal{O}(\gamma^2)\,,
\label{Xin}
\\
&&
\widehat{X}_{\mu\nu}^{\ \ \rho\sigma}=(X^{-1})_{\mu\nu}^{\ \ \rho\sigma}+\mathcal{O}(\alpha_2^2)+\mathcal{O}(\gamma^2)\,.
\ffa
It indicates that the self-dual is violated for the theory (\ref{action-A}).
But there is a duality between the actions (\ref{action-A}) and (\ref{ac-SB}) with the change of the sign of $\alpha_2$ or $\gamma$.

\begin{figure}
\center{
\includegraphics[scale=0.4]{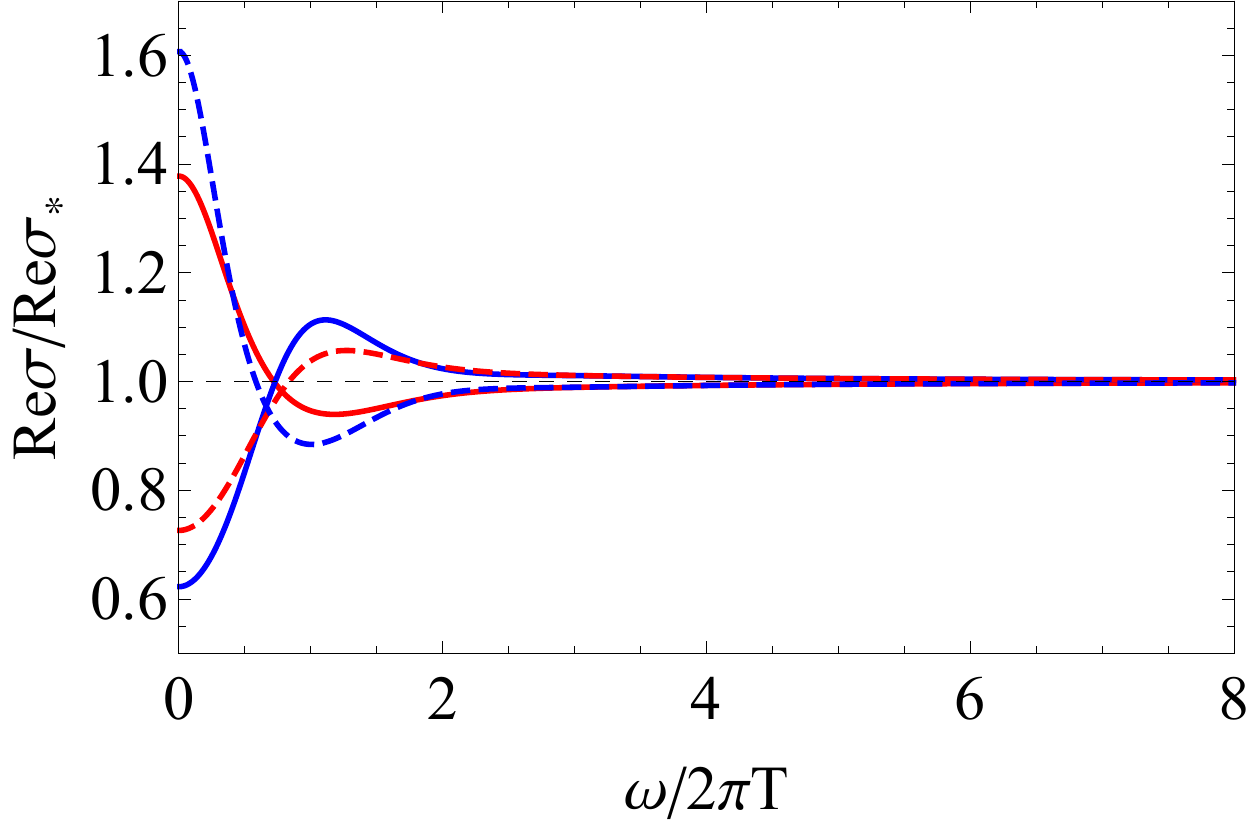}\ \hspace{0.1cm}
\includegraphics[scale=0.4]{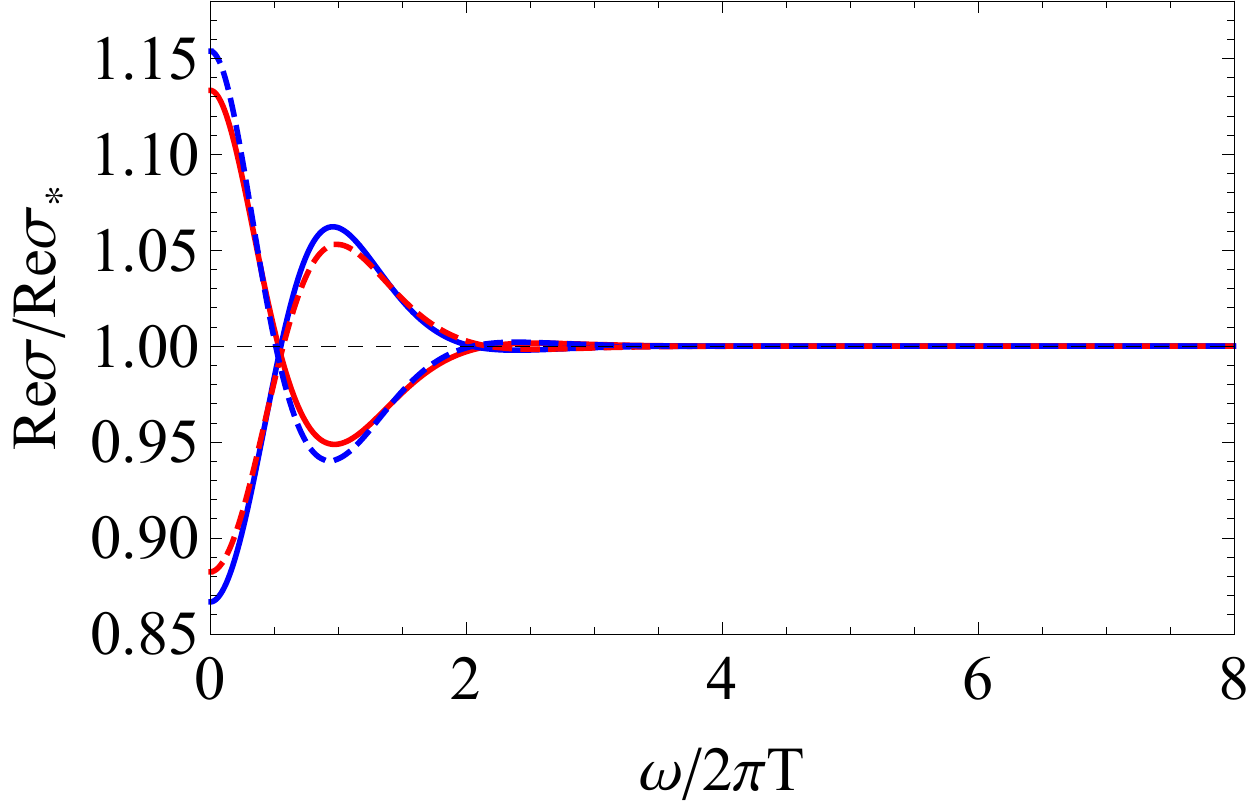}\ \hspace{0.1cm}
\includegraphics[scale=0.4]{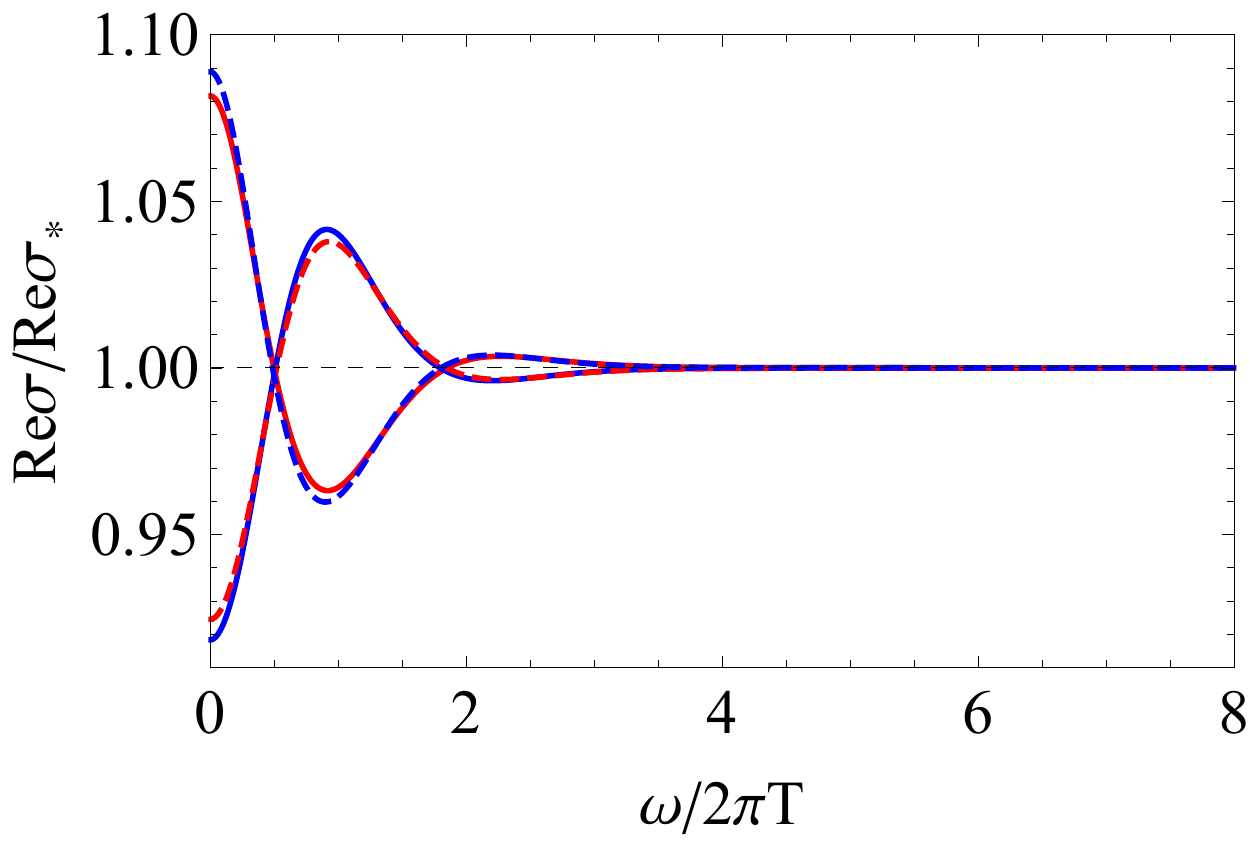}\ \hspace{0.1cm}\ \\
\includegraphics[scale=0.4]{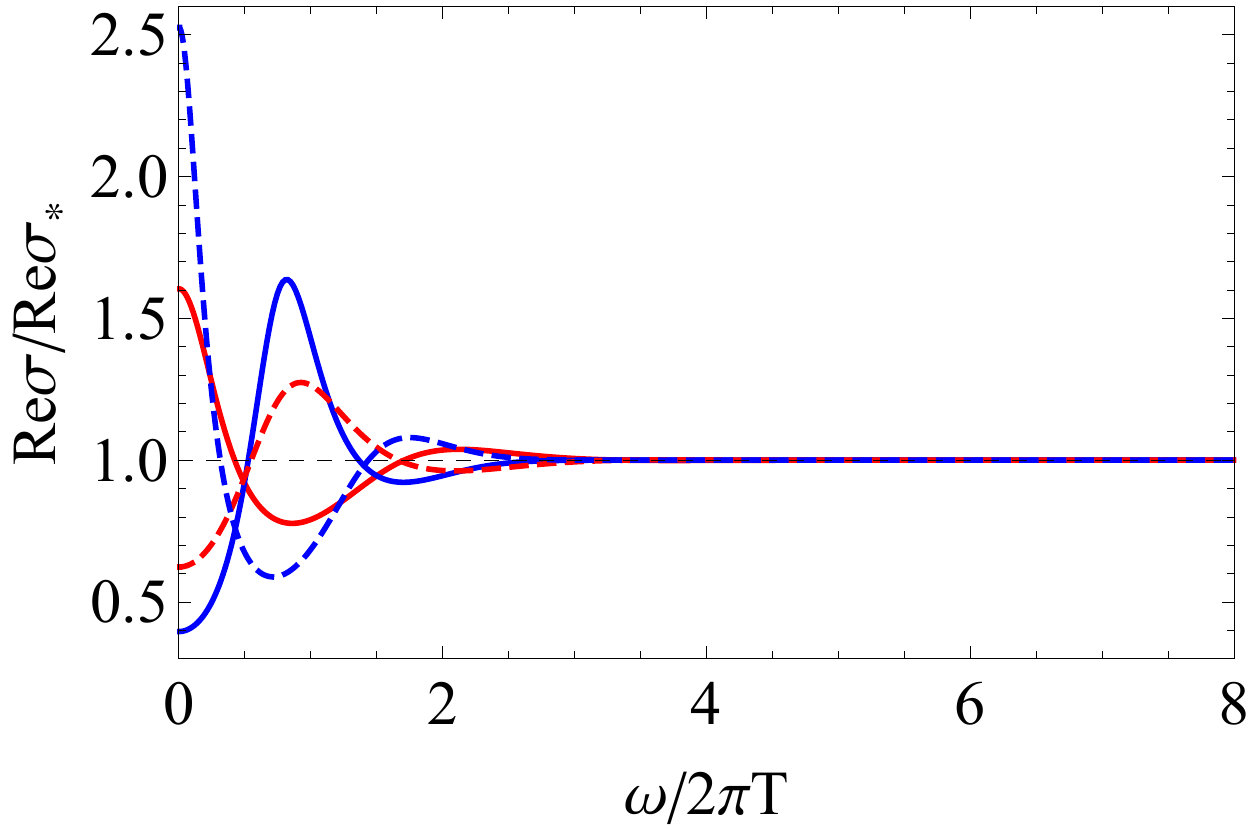}\ \hspace{0.1cm}
\includegraphics[scale=0.4]{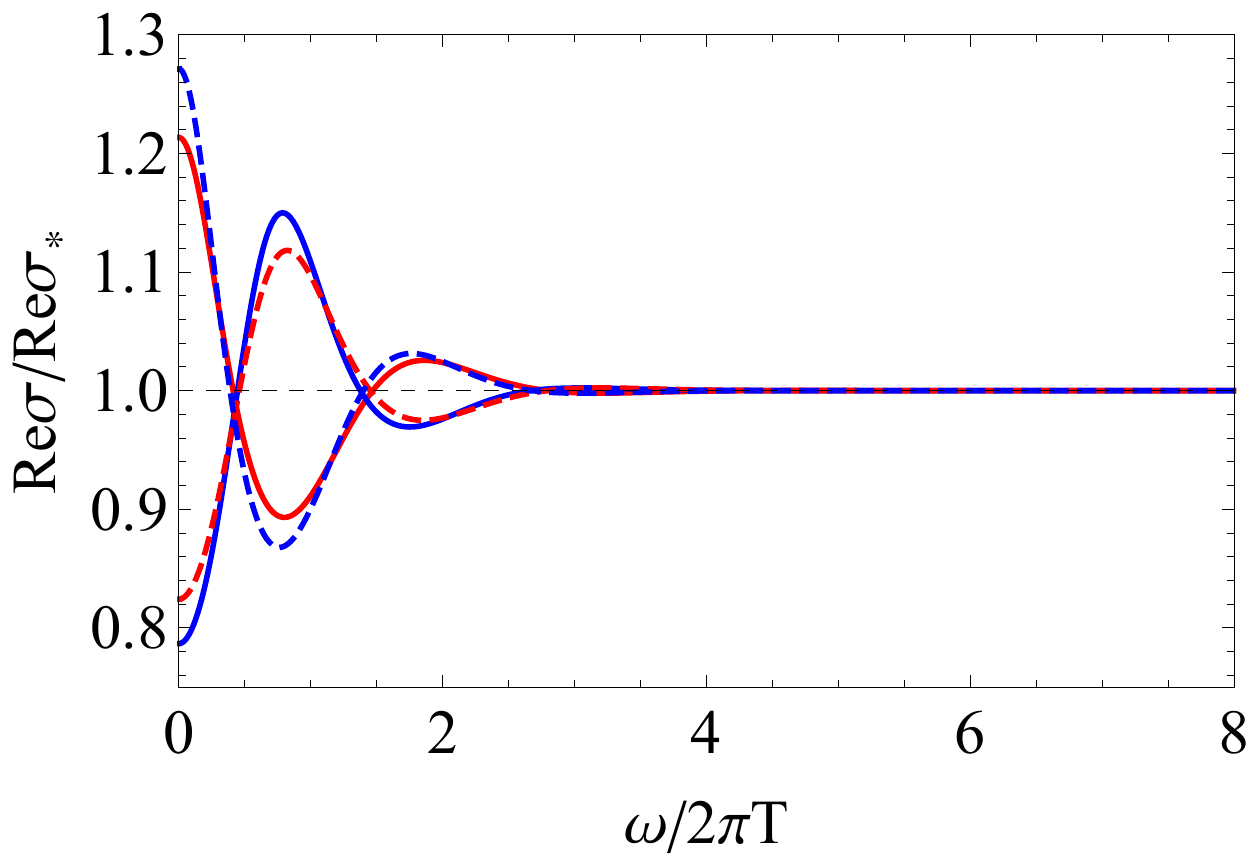}\ \hspace{0.1cm}
\includegraphics[scale=0.4]{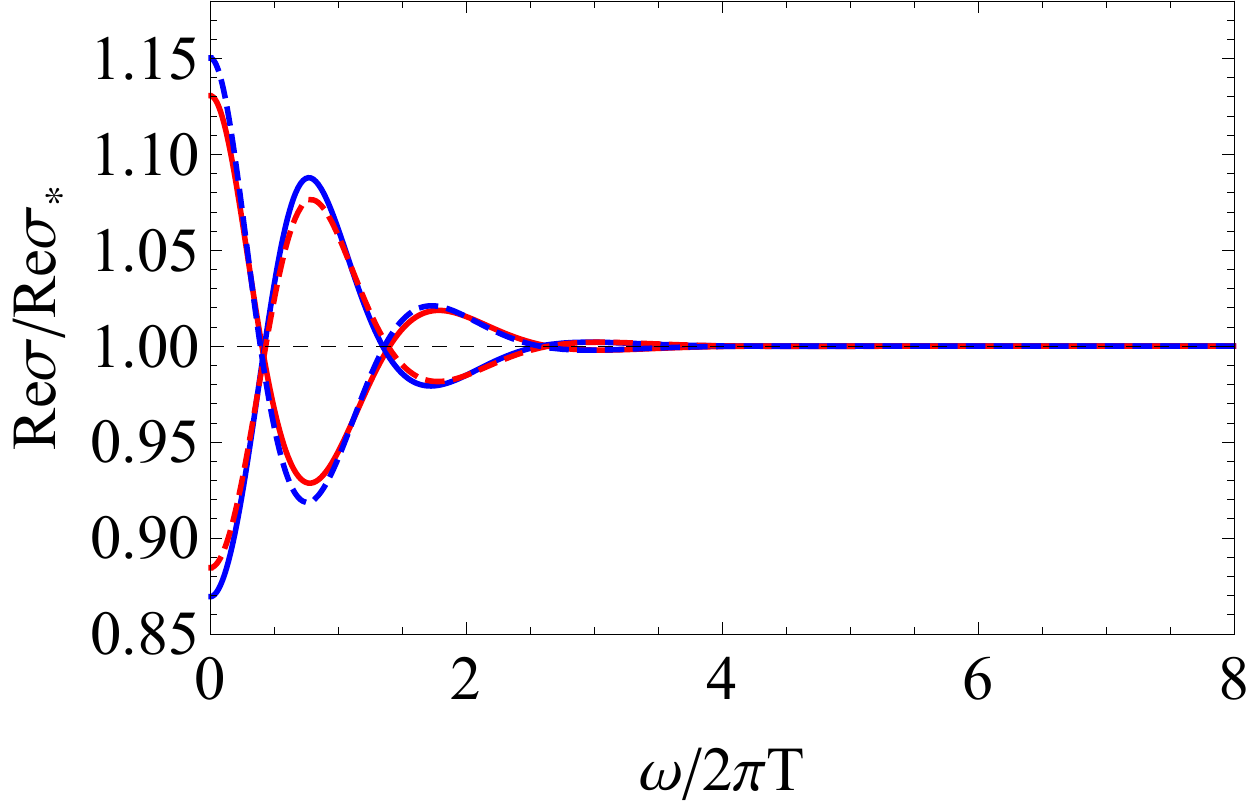}\ \hspace{0.1cm}\ \\
\caption{\label{em} The conductivity $\sigma(\omega)$ for various values of $\alpha_2$, $\gamma$ and $\Delta$.
The plots above are for $\alpha_2=\pm 0.2$, $\gamma=0$ and the ones below for $\gamma=\pm 0.08$, $\alpha_2=0$.
The plots from left to right are for $\Delta=1.5, 3, 4$, respectively.
The solid curves are the conductivity of the original EM theory (red for positive $\alpha_2$ or $\gamma$ and blue for negative $\alpha_2$ or $\gamma$).
While the dashed curves display the conductivity of the EM dual theory for the same value of $\alpha_2$ or $\gamma$.
Here we have set $\alpha_1=0.5$.
}}
\end{figure}

FIG.\ref{em} exhibits the conductivity $\sigma(\omega)$ of the bulk EM theory and its dual EM theory
for various values of $\alpha_2$, $\gamma$ and $\Delta$.
An oppositive picture appears in the dual EM theory.
That is to say, a peak at small frequency in optical conductivity occurs for $\alpha_2=-0.2$ or $\gamma=-0.08$,
while a dip displays for $\alpha_2=0.2$ or $\gamma=0.08$.
Since the EM duality in bulk, which corresponds to the particle-vortex duality in boundary theory,
with the change of the sign of $\alpha_2$ or $\gamma$, the conductivity is approximately same.
We notice that the largest discrepancies of the particle-vortex duality appear at the low frequency
and as expected, the particle-vortex duality holds more well for the irrelevant operator than for the relevant operator.
It is because the conductivity probes the geometry near the horizon
and the amplitude of the scalar field becomes larger near the horizon (see FIG.\ref{phivsu}).
Moreover, the amplitude of the scalar field near the horizon is more evident for the relevant operator than for the irrelevant operator (see FIG.\ref{phivsu}).

\section{Conclusions and discussions}

References \cite{Myers:2016wsu,Lucas:2017dqa} provide a natural mechanism to study the QC dynamics for a wide range of conformal dimensions $\Delta$.
We try to study the universality and the speciality of the dynamics of this QC system by introducing a higher derivative coupling term between the gauge field and the Weyl tensor.
We observe two key features of our present model, which are
\begin{itemize}
  \item To fit the QMC data for the $O(2)$ QCP, a rescaling for the Euclidean frequency by a constant is needed.
  It is different from the QC dynamics studied in \cite{Myers:2016wsu,Lucas:2017dqa} and seems to be a common characteristic of the higher derivative electrodynamics.
  \item Both the Drude-like peak at low frequency and the pronounced peak simultaneously emerge.
  In particular, they are more evident for the relevant operators than for the irrelevant operators.
  This pronounced peak is associated with the coupling between the nontrivial scalar field profile and the Weyl tensor.
\end{itemize}
In addition, we also further confirm that the conductivity for the $O(2)$ QCP is particle-like but not vortex-like
by studying the conductivity for Euclidean frequency.

Finally, we also briefly discuss the EM duality.
We find that the largest discrepancies of the particle-vortex duality in the boundary theory appear at the low frequency
and the particle-vortex duality holds more well for the irrelevant operator than for relevant operator.
It is because the amplitude of the scalar field becomes larger near the horizon and
the amplitude of the scalar field near the horizon is more evident for the relevant operator and for the irrelevant operator.

There are lots of open questions deserving further exploration.
\begin{itemize}
  \item First of all, it is important to study the conductivity at frequencies much greater than the temperature.
  We expect the contribution in the high-frequency expansion from higher derivative coupling.
  \item It is interesting and important to study the response and the dispersion of the quasinormal modes at full momentum and energy spaces.
  \item We can also study the holographic superconductivity based on this holographic framework.
  \item It is also important to further study the response away from QCP from higher derivative electrodynamics.
\end{itemize}
We plan to explore these questions and publish our results in the near future.

\begin{acknowledgments}

This work is supported by the Natural Science Foundation of China under
Grant Nos. 11775036, and by Natural Science Foundation of Liaoning Province under
Grant No.201602013.

\end{acknowledgments}

\end{document}